\documentclass{article}
\usepackage{PRIMEarxiv}

\usepackage[utf8]{inputenc} 
\usepackage[T1]{fontenc}    
\usepackage{hyperref}       
\usepackage{url}            
\usepackage{booktabs}       
\usepackage{amsfonts}       
\usepackage{nicefrac}       
\usepackage{microtype}      
\usepackage{lipsum}
\usepackage{fancyhdr}       
\usepackage{graphicx}       
\graphicspath{{media/}}     

\title{Modeling cognitive load as a self-supervised brain rate with electroencephalography and deep learning
}

\author{
  Luca Longo \\
  Artificial Intelligence and Cognitive Load Research Lab \\
  School of Computer Science, The Applied Intelligence Research Centre, \\
  Technological University Dublin \\
  Dublin, Republic of Ireland\\
  \texttt{\{luca.longo@tudublin.ie}\} \\
}

\usepackage{tikz}
\usetikzlibrary{positioning}
\usetikzlibrary{3d} 
\usetikzlibrary{arrows,shapes,automata,backgrounds,patterns,positioning,tikzmark,calc,chains,trees}
\usepackage[]{tikz-3dplot}
\usepackage{pgfplots}
\pgfplotsset{compat=1.10}

\usepackage{multicol}
\usepackage[T1]{fontenc}
\usepackage{times}%
\usepackage{amsmath}
\usepackage{enumitem}
\usepackage{dcolumn}
\usepackage{graphics}
\usepackage{amssymb}
\usepackage{graphicx}
\usepackage{makecell}
\usepackage{mathtools}
\usepackage{float}
\usepackage{array}
\newcolumntype{P}[1]{>{\centering\arraybackslash}p{#1}}
\usepackage{}
\usepackage{bm}

\usepackage{multirow}
\usepackage{lipsum}
\usetikzlibrary{positioning}
\usetikzlibrary{patterns}
\usepackage{schemata}
\usepackage{enumerate}
\usepackage{subcaption}
\usepackage{pifont}
\usepackage{xcolor}    
\pgfplotsset{compat=1.9}
\usepackage{tikz}
\usepackage{booktabs}
\usepackage{url}

\begin{document}

\maketitle

\begin{abstract}
The principal reason for measuring mental workload is to quantify the cognitive cost of performing tasks to predict human performance.
Unfortunately, a method for assessing mental workload that has general applicability does not exist yet. 
This is due to the abundance of intuitions and several operational definitions from various fields that disagree about the sources or workload, its attributes, the mechanisms to aggregate these into a general model and their impact on human performance.
This research built upon these issues and presents a novel method for mental workload modelling from EEG data employing Deep Learning.
This method is self-supervised, employing a continuous brain rate, an index of cognitive activation, and does not require human declarative knowledge.
The aim is to induce models automatically from data, supporting replicability, generalisability and applicability across fields and contexts.
This specific method is a convolutional recurrent neural network trainable with spatially preserving spectral topographic head-maps from EEG data, aimed at fitting a novel  brain rate variable.
Findings demonstrate the capacity of the convolutional layers to learn meaningful high-level representations from EEG data since within-subject models had, on average, a test Mean Absolute Percentage Error of around 11\%.
The addition of a Long-Short Term Memory layer for handling sequences of high-level representations was not significant, although it did improve their accuracy. 
These findings point to the existence of quasi-stable blocks of automatically learnt high-level representations of cognitive activation because they can be induced through convolution and seem not to be dependent on each other over time, intuitively matching the non-stationary nature of brain responses. 
Additionally, across-subject models, induced with data from an increasing number of participants, thus trained with data containing more variability, obtained a similar accuracy to the within-subject models.
This highlights the potential generalisability of the induced high-level representations across people, suggesting the existence of subject-independent cognitive activation patterns.
This research contributes to the body of knowledge by providing scholars with a novel computational method for mental workload modelling that aims to be generally applicable and does not rely on ad-hoc human crafted models.

\end{abstract}

\section{Introduction}
The explosion of home-working and online interactions, the pervasive uses of technologies in daily activities and many working environments impose ever more mental workload upon operators and less physical load.
The literature on the construct of mental workload (MWL) or, often interchangeably referred to as cognitive load (CL), has been vast and in constant evolution for the last half-century. 
Note that cognitive load and mental workload might differ in little aspects, according to authors working in different fields, such as psychology, neuroscience or education. However, to my knowledge, no clear evidence has been found to address these differences formally. Thus they are interchangeably used in the remainder of this article \cite{longo2022human}.
The principal reason for measuring workload is to quantify the mental cost of performing tasks to predict human performance \cite{wickens2017mental}. In turn, prediction of performance can be used for designing interfaces, interactive technologies \cite{longo2015designing}, and information-processing activities \cite{OrruL19} optimally aligned to the well-known human mental limited capacities \cite{miller1956magical}. 
Despite 50 years of effort, research on MWL has not been able to make major advances yet \cite{hart2006nasa,young2001mental,paas2003cognitive} failing at providing a model of understanding \cite{young2015state,charles2019measuring,van2018understanding,HancockLongo2021}. Guess intuitions and several operational definitions from various fields have proliferated \cite{longo2018reliability,HART1988139}. Still, these disagree about the MWL sources, their attributes, the mechanisms to aggregate these together and their impact on human performance \cite{young2015state}. Identifying these sources, attributes, and mechanisms and how they impinge on human performance are all open fundamental research problems. For instance, some researchers have considered task-specific attributes \cite{wickens2020processing} while others chose a combination of task and user-specific attributes \cite{hart2006nasa}. Primary researchers have employed self-reporting measurements \cite{HART1988139} or a combination of psychophysiological techniques \cite{Brookhuis2001}. However, MWL is also influenced by the environment in which a human performs a task \cite{Vidulich2012}. \\


Currently, the literature on mental workload includes a plethora of hand-crafted knowledge-driven models grounded in different theories, employing different attributes and different strategies for aggregating these into indexes of workload, limiting their comparison \cite{cain2007review,longo2015defeasible,van2018understanding, longo2022human}. This makes cognitive load a \emph{knowledge-dependent construct}.
This is also supported by the fact that cognitive load has been mainly investigated in the fields of ergonomics and psychology \cite{HART1988139,young2015state} with several applications in the aviation \cite{hart2006nasa}, automobile \cite{Brookhuis2001} and manufacturing industries \cite{bommer2018theoretical}.
In these fields, investigations are mainly conducted in labs and highly controlled settings, making cognitive load a \emph{field-dependent construct}.
Past research has had a tendency to focus on complex safety-critical systems \cite{charles2019measuring} with many applications in the transportation \cite{arico2016adaptive,borghini2014measuring}, nuclear and manufacturing industries \cite{hart2006nasa,Brookhuis2001}, making mental workload an \emph{application-driven construct}. However, researchers have claimed the need for models of cognitive load in other ecological settings with real-world activities \cite{young2015state,young2001mental,paas2003cognitive,burns2018understanding}. 
The vast majority of existing knowledge-dependent, field-dependent, and application-driven models aggregate attributes, believed to influence workload, in a linear fashion \cite{hart2006nasa}, or assume stationarity within a task, neglecting temporal dynamics \cite{charles2019measuring}, making cognitive load a \emph{static construct}. 
Additionally, these models are largely built by fitting or correlating to some ad-hoc indicator of human performance. This is either explicitly achieved by applying self-reporting techniques and correlating to subjective responses from experimental participants, or based on fitting human responses grouped by tasks of varying demands, often ad-hoc and subjectively defined. 
This largely complicates research efforts attempted at modelling mental workload and increasing the generalisability of models because they are highly constrained on those subjective design choices from modellers that highly differ across experiments, disciplines and contexts.
The aforementioned state of the art in cognitive load modelling has led to many definitions of workload \cite{cain2007review,xie2000review,johannsen1979workload,longo2022human} and the formation of ad-hoc, knowledge-dependent, field-dependent, application-driven and static models with little chance of reconciliation \cite{charles2019measuring}. Because of this, despite 50 years and more of research, the construct of workload is still ill-defined \cite{cain2007review,young2015state, paas2003cognitive,longo2015defeasible, charles2019measuring, longo2022human}. \\


The goal of this research is to tackle the above issues and design a model of cognitive load that has wider applicability, facilitating comparison across studies, that is less constrained to the context of application, that is not static, that does not require any explicit ground truth, and that minimizes experimental design-choices of researchers.
To achieve this goal, this research proposes to apply modern Deep Learning methods to avoid incrementally extending current knowledge-driven approaches and supporting automatic learning of salient features for cognitive load and their non-linear inner-relationship from data. 
Additionally, this research focuses on neurophysiological data collected in ecological settings and daily real-world activities not traditionally considered in cognitive load research. 
In detail, electroencephalography is employed for such a purpose.
Experiments will be focused on simultaneously taking advantage of the temporal, spatial and spectral properties of physiological EEG data without making any assumption on the linearity of cognitive load, supporting the automated extraction of salient features and representations and their inner relationships with no explicit declarative knowledge from designers. This will allow moving beyond the knowledge-driven research approaches that have produced hand-crafted deductive knowledge and have dominated the research landscape on mental workload for the last 50 years. Also, without resorting to self-reporting subjective perceptions or task-performance measures but only manipulating physiological EEG data, it represents a more objective method for modelling cognitive load. 
Eventually, the proposed computational method does not require explicit ground truth for mental workload. Instead, a self-supervised brain rate generated from data is proposed, supporting the development of a model of cognitive load that potentially has a higher degree of replicability and applicability.\\

The remainder of this article is structured as follows.
Section \ref{sec:design}, introduces the design of a self-supervised mental workload model based on a brain rate, an index of cognitive activation, trained with deep learning techniques that are expected to identify recurrent patterns while fitting such a rate. 
Section \ref{sec:results} presents the results of the experiment, followed by a discussion in section \ref{sec:discussion} and the identification of future research improvements.

\section{State of the art in cognitive load modeling}
The literature on cognitive load is vast, and recent work has attempted to collate the great amount of information surrounding this construct \cite{longo2022human}. 
This section thus is mainly devoted to reviewing related work on the application of Electroencephalography to the problem of cognitive load modelling and not performing another wide general review of this construct.
\emph{Electroencephalography} (EEG) is a technique for the direct assessment of brain activity via electrodes placed on the scalp and, as a consequence, the inference of objective neuro-metrics of human mental activation and mental states \cite{richer2018real}.
The advantages behind the application of EEG data for cognitive load modelling are represented by its low invasiveness, when compared to neuroimaging methods such as fMRI \cite{lemieux1997recording}, its wider applicability in ecological settings, thanks also to its high portability \cite{xu2018review,casson2010wearable} and financial affordability \cite{mullen2015real}, and its high temporal resolution \cite{burle2015spatial}. 
Unfortunately, EEG-based cognitive load modelling methods must consider several technical issues. Firstly, variation in EEG signals exists mainly because of the slight differences in cortical mappings and brain functioning of subjects, leading to differences in spatial, spectral and temporal patterns or due to imperfect fitting of the EEG cap on heads of different shapes and sizes. Therefore, a key challenge in successfully recognizing mental states from EEG data is to create a model that is \emph{robust to deformation and translation of signal in space, frequency, and time} due to inter and intra-subject differences and to the protocols or methods employed in signal acquisition. 
Fortunately, advances in machine learning \cite{jordan2015machine} and particularly in deep learning methods \cite{lecun2015deep} have proven useful for learning models from EEG data \cite{craik2019deep}. The advantage of these \emph{data-driven deep-learning methods} is that they support the automatic extraction of meaningful high-level representations from complex, non-linear data \cite{GomezLongo2022}, they can lead to the creation of learning architectures that have wider applicability, supporting replicability of experimental research, and are flexible enough to be adapted and extended, eventually supporting advances and research progresses.
However, applications of Deep Learning methods with EEG data have barely attempted to jointly preserve the structure of EEG signals within space, frequency and time. Most studies have focused on spatio-temporal learning \cite{tran2015learning}, time-frequency learning \cite{boashash2016automatic} or spatial-frequency learning \cite{ang2012filter}. Therefore, a challenge is to inductively learn a model capable of exploiting the spatio-temporal and frequency-based properties of EEG data.\\

The literature on cognitive load modelling with EEG and deep learning is recent, not vast and highly scattered \cite{saha2018classification, jimenez2020custom, qayyum2018classification, bashivan2015single, liu2017convolutional, jiao2018deep, xiong2020pattern, cabanero2019analysis, yin2017cross}. Most of these models are supervised, which means they require a form of ground truth, usually in task-based categories or task-performance measures. Unfortunately, there is no agreement among researchers on how to form such categories systematically. This limits comparisons across studies because, on the one hand, some scholars might focus on building a model for classifying low or high levels of task load for relatively simple tasks. On the other hand, others might focus, for example, on building models for assessing low, medium or high load of complex tasks. In other words, these models are context-dependent, and they learn high-level features from EEG data focused on fitting these application-specific target classes. 
Therefore they cannot be meaningfully used across studies, limiting their generalisability. 
Some recent work focused on applying unsupervised learning techniques such as auto-encoders to automatically learn relevant latent representations from EEG data in an unsupervised fashion or aimed at automatically reducing the presence of noise in the data itself \cite{yang2019assessing, yin2019physiological}. However, these unique high-level representations are often used to learn a second model that, unfortunately, still often requires supervision, as the goal is to fit, as described earlier, categories of task load, these being the independent feature subjectively defined by researchers. 
State-of-the-art models manipulating EEG data often rely on frequency bands, such as the alpha or theta rhythms, deemed the alphabet for brain functions and mental state extraction. These have been individually used as cognitive load indicators \cite{stipacek2003sensitivity,castro2020validating}, or aggregated together  \cite{chang2016yet,holm2009estimating,borghini2014measuring,RaufiLongo2022} because they have been shown to be sensitive to task difficulty manipulation, task engagement or memory load \cite{gevins2003neurophysiological,antonenko2010using}.
However, these approaches often discard some EEG bands in favour of other bands.

\section{Design and methods} \label{sec:design}
A novel method is proposed to tackle the issues in modelling cognitive load, as discussed in the previous sections, followed by an empirical study to validate such a method. 
Contrary to all the existing methods of cognitive load modeling, the method proposed here is \emph{self-supervised} \cite{jing2020self,banville2021uncovering}.
Self-supervision is an approach that autonomously learns from the data itself, and that is in the middle between supervised and unsupervised learning methods within the discipline of artificial intelligence. 
It is not fully supervised because it does not require ground truth (an independent variable to fit), usually as a form of declarative knowledge.
It is also not fully unsupervised because it is not used for discovering patterns in the EEG data that need to be subsequently labelled and categorised with human intervention. 
Rather, self-supervision refers to the fact that the ground truth is generated by some automatic methods applied to the available data itself. 
Subsequently, some supervised machine learning algorithm uses this ground truth as supervisory data to train a model. 
In other words, self-supervised machine learning can be seen as an autonomous form of supervised learning because it does not require explicit human declarative knowledge. \\

\begin{figure}[ht]
\includegraphics[scale=0.36]{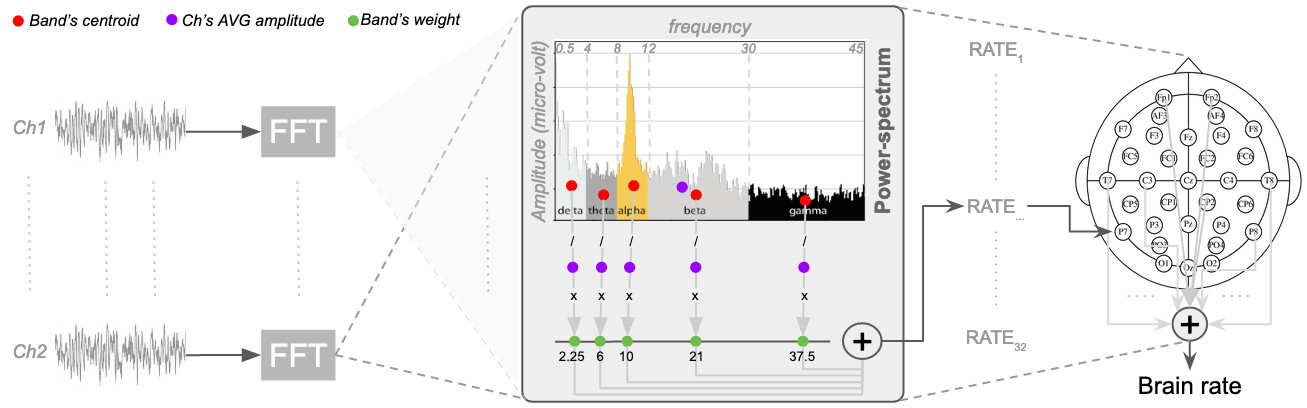}
\caption{Diagrammatic illustration of the computation of the mean frequency of brain oscillations weighted over the EEG bands of potential (power) spectrum for each channel and their final aggregation towards a brain rate.}
\label{fig:brainRateComputation}
\end{figure}

Analogously to blood pressure and heart rate, seen as standard preliminary indicators of general bodily activation, a brain rate is proposed as an indicator of mental activation, and then used in this research as an indicator of cognitive load.
In contrast to the approaches that suppress or elevates some EEG band, as described in the previous section, the proposal is to fully use them, reasonably assuming that, whenever some band is modulated, the others are influenced too \cite{ferri2008functional}. Analogously to the computations for the centre of gravity or the mean energy of a physical system \cite{landi2002properties}, a spectrum-weighted frequency rate across the five canonical EEG bands (delta, theta, alpha, beta, gamma) is proposed \cite{pop2005spectrum}, here on referred to as the \emph{brain rate} (BR). This is the sum of the mean frequency of brain oscillations weighted over the EEG bands of the potential (power) spectrum for each channel, as illustrated in figure \ref{fig:brainRateComputation}).
Formally:
$$
BR=\sum_{ch=1}^{n} \sum_{b=1}^{5} f_b \cdot P(b,ch)
$$
where $b$ is the index denoting the frequency band (for delta $b=1$, theta $b=2$, alpha $b=3$, beta $b=4$, gamma $b=5$), 
$ch$ is the index denoting a specific EEG channel,
$f_{b}$ is the weight associated with frequency band $b$, which is the mean frequency of each EEG band. Setting the boundaries for each band as delta=\{0.5-4 hz\}, theta=\{4-8 hz\}, alpha=\{8-12 hz\}, beta=\{12-30 hz\} and gamma=\{30-45 hz\}, then $f_1=2.25, f_2=6, f_3=10, f_4=21, f_5=37.5$ (Figure \ref{fig:brainRateComputation}).
$P(b,ch)$ is the mean amplitude of the electrical potential for band $b$ of a channel $ch$ over the mean of all its amplitudes:
$$
P(b,ch) = \frac{avg_b(FFT_{ch})}{avg(FFT_{ch})} 
$$
with 
$FFT_{ch}$ is the vector containing the amplitudes of the fast-Fourier transformed channel $ch$,
$avg_b$ is the average (centroid) of only the amplitudes within the frequency band $b$.
Note that $f_b$ is in hertz, and $P(b,ch)$ is in micro-volt, with the brain rate $BR$ in hertz.  By keeping the length of an EEG segment relatively short, in the order of seconds, then this rate can be used as a pseudo-real-time measure of cognitive load, since it is the mean activation of the brain response, as registered all over the scalp. Pseudo-real-time is because this rate is computed over a window of EEG data rather that a single point in time. This is also dictated by the fact that the Fourier transformation requires some data collected over time to produce meaningful translations in the frequency domain.\\

\begin{figure*}
\centering
    \includegraphics[scale=0.25]{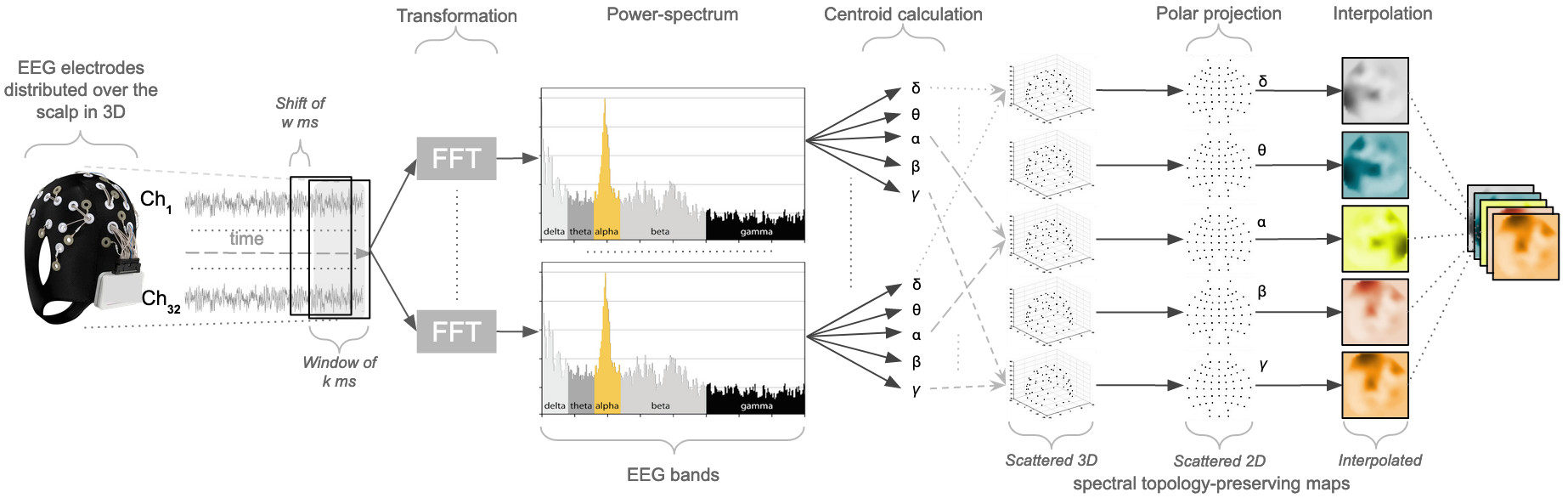}
\caption{Processing pipeline for producing topology-preserving head maps from windows of EEG data. I) The electrodes distributed over the scalp in a 3D space produce neural signals continuously over time II) these are segmented into windows  III) for each signal in a window, fast-Fourier transformation is applied to obtain information in the power spectrum IV) each power-spectrum is divided into the five EEG bands (delta, theta, alpha, beta, gamma) V) the centroid of the frequency amplitudes for each band is computed VI) all the centroids are positioned in a 3D space to produce a scattered head map, one for each EEG band VII) polar projection is applied to each scattered map to produce 2D head maps VIII) each 2D map is interpolated IX) the 5 2D maps, one for each EEG band are aggregated into a tensor.}
    \label{fig:processingTopographicMaps}
\end{figure*}

One common problem within neuroscience, in general, and for the specific technical challenge of creating a robust model of cognitive load, in particular, is the limited availability of EEG data. This is often due to the difficulties in recruiting participants, or faulty recordings, or the presence of various artefacts in the EEG signal, leading researchers to discard significant portions of collected data. 
Unfortunately, when employing machine learning methods, in general and deep learning methods in particular, limited training data might often not benefit a robust model formation.
For these reasons, this work proposes to use a sliding-window technique \cite{ryang2016high}.
The available EEG data are segmented into windows of $k$ seconds, shifted by $w$ milliseconds. For each window, a pre-processing pipeline has been designed for producing 2D spatial-spectral preserving images, as summarised in figure \ref{fig:processingTopographicMaps}. Fast Fourier transformation is run for each EEG channel in each window, obtaining a power spectrum in the frequency domain. For each spectrum, the five EEG bands  (delta, theta, alpha, beta, gamma) are defined by employing the same boundaries used to compute the brain rate. For each band, the centroid (geometric centre) is computed, which equates to the arithmetic mean of all the power values within that band. For a given band, all the computed centroids, one for each channel, are positioned in a 3-dimensional space, following the coordinates of each electrode position on the scalp, forming a scattered 3D spectral topology-preserving map. Azimuthal Equidistant Projection (polar) is subsequently used to transform this map into a scattered 2D map, preserving the relative distance between adjacent electrodes. Eventually, the Clough-Tocher method \cite{mann1999cubic} is applied to fill the scattered 2D maps by estimating the values in-between the electrode over a new interpolated map, an image of $32x32$.
The aggregation of the five $32x32$ maps, one for each EEG band, creates a tensor of $32x32x5$.
The sequence of these tensors can be seen as an `EEG movie', a stream of data over time in the frequency domain that preserves information in space. This stream can then be processed with  deep learning methods, inspired by state-of-the-art video classification methods for spatio-temporal feature learning \cite{yue2015beyond,wang2018appearance}.\\

\begin{figure*}
\centering
\includegraphics[scale=0.41]{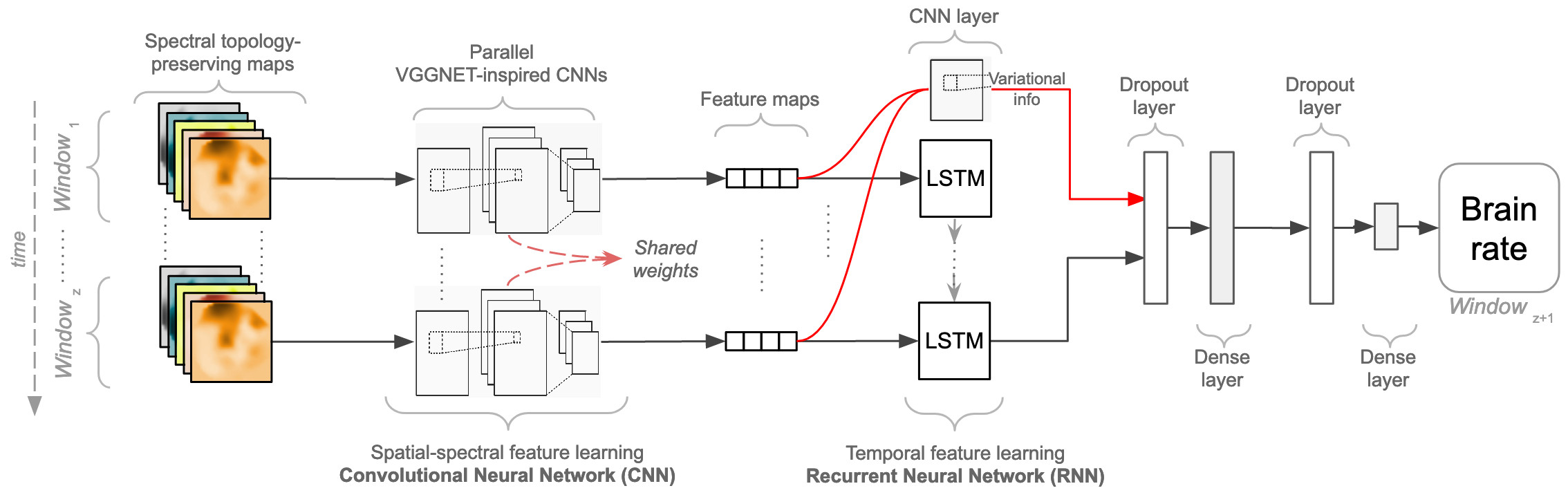}
\caption{Self-supervised Convolutional-recurrent deep neural network for spatio-temporal learning with spectral topology-preserving head maps and a brain rate.}
\label{fig:neuralnetwork}
\end{figure*}

The aforementioned justifications and design choices have led to the design of a novel self-supervised convolutional, recurrent deep neural network trained to fit the brain rate introduced above.
The proposed architecture, as depicted in figure \ref{fig:neuralnetwork}, is built upon a first part, the Convolutional Network (CNN), due to its ability to learn robust compressed representations of EEG data, and upon a second part, the Recurrent Network (RNN) to account for temporal variations. 
From a higher perspective, the overall architecture contains $z$ parallel convolutional networks with shared weights, which are useful for representational learning. Their outputs, high-level representations referred to as feature maps, are concatenated into a sequence of length $z$, respecting their time order. This sequence is subsequently injected into a recurrent network composed of Long Short-term Memory units (LSTM) aimed at temporal feature learning.  The feature maps, the output of each CNN parallel network, are injected into a final convolutional one-dimensional layer, and along with the output of the last LSTM unit, they are used to fit the brain rate extracted from the $z+1$ EEG window (hence self-supervision).\\

\begin{figure*}
\centering
\includegraphics[scale=0.34]{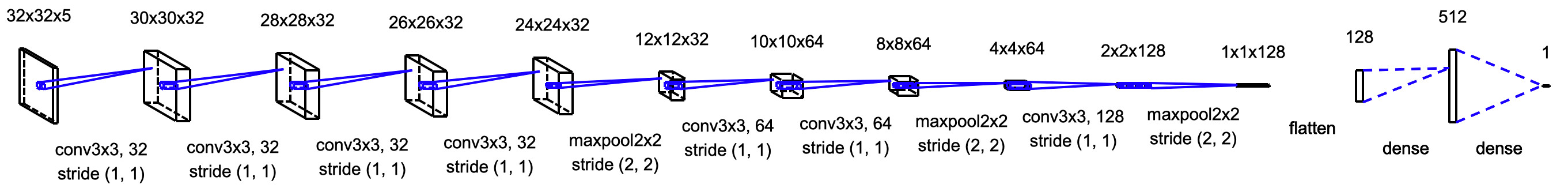}
\caption{Single VGGNET-inspired Convolutional Neural Network (CNN) architecture for feature maps learning with spectral topology-preserving head-maps with brain rate as a target feature.}
\label{fig:cnn}
\end{figure*}

In more detail, the CNN architecture was inspired by the VGG-NET architecture designed and used in the Imagenet classification challenge \cite{simonyan2014very,russakovsky2015imagenet}. 
In detail, this network, as depicted in figure \ref{fig:cnn}, is composed of $7$ stacked convolutional layers with small receptive fields of size $3x3$ and stride of $1x1$ pixel, with Rectified Linear Unit (ReLU) selected as the activation function. 
To preserve the spatial resolution of each of the $32x32x5$ topology-preserving spectral maps of each convolutional block, each layer's inputs are padded with $1$ pixel.
Each stacked block of convolutional layers is followed by a max-pooling layer over a $2x2$ window with a stride of $2x2$ pixels.
The number of kernels in each convolutional block doubles for every consecutive block, expecting to create effective receptive fields of higher dimensions while requiring fewer parameters \cite{simonyan2014very}. 
In summary, this network contains $4$ consecutive 2D CNN layers with $32$ filters, each with a kernel size of $3x3$, a stride of $1x1$ and no padding (`valid' padding), followed by a max pooling layer with a stride size of $2x2$ and zero-padding (`same' padding, results in padding with zeros evenly to the left/right or up/down of the input). This block is followed by another one containing two 2D-CNN layers with $64$ filters, with a kernel size of $3x3$, a stride of $1x1$ and no padding (valid padding), followed by a max pooling layer with a stride size of $2x2$ and zero-padding (same padding). Eventually, the last block contains a single 2D-CNN layer with $128$ filters,  with a kernel size of $3x3$, a stride of $1x1$ and no padding (valid padding), followed by a max pooling layer with a stride size of $2x2$ and zero-padding (same padding).
Since the nature of neural responses is dynamic over time, a suitable method for modelling the temporal evolution of brain activity is recurrent neural networks (RNNs).  Technically, Long Short-Term Memory (LSTM) appears to be an appropriate modelling choice \cite{hochreiter1997long}. It is a specific type of RNN that uses memory cells with internal memory, and gated inputs/outputs which
have led to the creation of models that are efficient in capturing long-term dependencies. The hidden layer function for LSTM is calculated by applying the following  equations:

$$i_t = \sigma \bigl( W_{xi^{x_t}} + W_{hi^{h_{t-1}}} + W_{ci^{c_{t-1}}}+ b_i \bigr)$$
$$
f_t = \sigma \bigl( W_{xf^{x_t}} + W_{hf^{h_{t-1}}} + W_{cf^{c_{t-1}}}+ b_f \bigr)
$$
$$
c_t = f_t c_{t-1} + i_t tanh \bigl( W_{xc^{x_t}} + W_{hc^{h_{t-1}}} + b_c \bigr)
$$
$$
o_t = \sigma \bigl( W_{xo^{x_t}} + W_{ho^{h_{t-1}}} + W_{{co}^{c_t}} + b_o \bigr)
$$
$$
h_t = o_t tanh(c_t)
$$
 $\sigma$ represents the logistic sigmoid function, $i$ as the input gate of the LSTM model, $f$ as its forget gate, $o$ as the output gate and $c$ as the cell activation vectors.
As shown in \cite{bashivan2016mental} where various trials were performed with EEG data, a reasonable number of LSTM units seems to be only one, with $128$ cells in it. This architecture was adopted to capture the temporal relationship of the feature maps obtained from each parallel CNN and shaped as a sequence of feature maps. 
However, only the output made by the LSTM after seeing the complete sequence of the feature maps produced by each parallel CNN was propagated to a fully connected layer.
This fully connected layer also gets the output of a CNN layer that receives the concatenation of the features maps computed by each of the parallel CNNs. This is because of the reasonable assumption that variations between these may contain additional information about the underlying mental state experienced by a subject. 
This is a single 2D-CNN layer containing $64$ filters with a stride of dimension $1x1$  with valid padding and ReLU as the activation function.
The output of this layer was concatenated to the output of the last LSTM, followed by a drop-out layer with a probability of $0.5$, and its output was injected to a dense layer with $512$ neurons and ReLU as an activation function.
Another dropout layer with a probability of $0.5$ followed, and a final dense layer with a linear activation function was devised for fitting the brain rate computed for the next window of EEG data following the sequence in time ($z+1$). 
Concerning the hypothesis that this study seeks to test, this is:\\

\begin{quote}
H: IF a convolutional-recurrent deep neural network architecture is trained with spatio-temporal spectral topology-preserving head maps, derived from multi-channel EEG data, to fit a brain rate, an index of cognitive activation, in a self-supervised fashion\\
THEN  within-subject and across-subjects models can be induced with low error rates, highlighting recurrent patterns of cognitive activation, and thus cognitive load.\\
\end{quote}

In order to test such a research hypothesis, data from a well-known dataset of EEG recording is employed, namely, the DEAP dataset \cite{koelstra2011deap}, as described in the following section.

\subsection{Dataset and pre-processing}
Electroencephalographic (EEG) data was recorded from 32 participants while watching 40 one-minute long excerpts of music videos \cite{koelstra2011deap}. These segments were carefully selected with maximum emotional content, following a procedure that considers valence and arousal as emotions.
The main rationale behind the selection of this dataset was the fact the data was recorded for a prolonged time, that means 1 minute, and not in the order of seconds,  as often the case for event-related potential studies. The reasonable assumption was that while cognitively processing excerpts of videos and perceiving different emotions, participants would have also experienced different levels of cognitive load \cite{plass2019four}. 
Cortical activity was recorded at 512 Hz using a Biosemi ActiveTwo system using 32 active AgCl electrodes placed according to the international 10-20 system,  with participants sitting 1 meter away from a 17-inch screen. A 5-second fixation cross was run before each video to act as a baseline. Participants watched two blocks of 20 videos each, separated by a short break. 
Other peripheral physiological signals and self-reports were recorded in the original experiment \cite{koelstra2011deap}. However,  only EEG data from the following 32 EEG channels were considered: Fp1, AF3, F3, F7, FC5, FC1, C3, T7, CP5, CP1, P3, P7, P03, O1, Oz, Pz, Fp2, AF4, Fz, F4, F8, FC6, FC2, Cz, C4, T8, CP6, CP2, P4, P8, PO4, O2. A pre-processing procedure using the EEGlab toolbox was applied to data, including i) downsampling to 128Hz ii) EOG artefact removal using a blind-source separation technique iii) band-pass filtering between 4.0Hz to 45Hz iv) common average referencing v) 3 seconds pre-trial baseline was kept. For further information, readers are referred to \cite{koelstra2011deap}.

\begin{figure}[ht]
\centering
\includegraphics[scale=0.50]{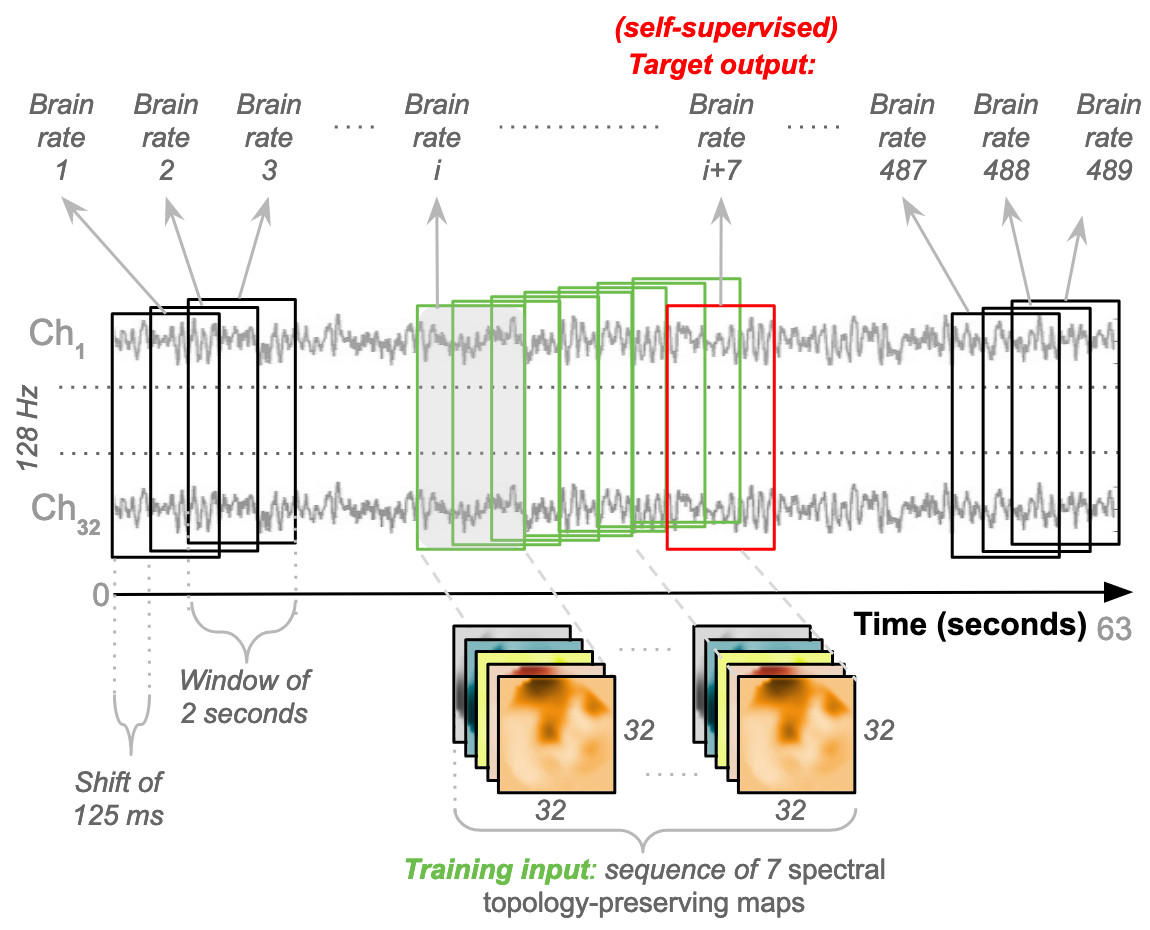}
\caption{Pipeline for generating sequences for the convolutional-recurrent neural network.}
\label{fig:pipelineTraining}
\end{figure}

\subsection{Training}\label{sec:training}
After the pre-processing pipeline is applied to selected EEG data, a new procedure (as depicted in Figure \ref{fig:pipelineTraining}) is designed and run to generate training instances for the specific convolutional/recurrent neural network described in the previous section.
Here, each video that participants watched lasted for 63 seconds (60 for the actual video and 3 for baseline). A time window of 2 seconds is set for producing spectral topology-preserving maps by applying the processing pipeline described in Figure \ref{fig:processingTopographicMaps}. 
This length is deemed short enough for producing a meaningful power spectrum that contains enough points well distributed across the five EEG bands. In detail, given a final sample rate of $128$ Hz, each window contains $256$ points ($128x2$) spread across the EEG bands for each channel. This means that each video contains 8064 points ($63x128=8064$). A sliding-window technique is applied across these points, and a shift of $125$ ms is used ($8$ points per second), which translates into a shift of 16 points ($128x0.125$). This generates $489$ windows of $2$ seconds ($63x8-16+1$) for each video in the dataset. The neural network designed in figure \ref{fig:neuralnetwork} is a specific convolutional-recurrent neural network accepting a sequence of windows.
As mentioned before, this sequence is set to $z=7$ windows, equating to $14$ seconds of neural activity. This is believed to be short enough for the expectation of detecting some variations in cognitive load, and not too long for hampering the automatic learning of temporal dependencies across points. Each of these sequences represents a training input instance. Thus $482$ of these instances (sequences) were produced for each video ($489-7$). 
As previously mentioned, the designed architecture is a specific self-supervised many-to-one network. The target output is the brain rate computed for the subsequent window outside the sequence, next in time (the $8th$). 
The goal is to learn this rate from past information, which in other words, is the estimation of a brain rate from the neural activity of the previous $14$ seconds ($7x2$).\\

\begin{table*}[ht]
    \centering
    \begin{tabular}{| l| l| c| c| c | c| c| }
    \hline
         \multirow{2}{*}{Models}  & \multirow{2}{*}{Type}  & \multicolumn{4}{c|}{Instances (training sequences)} & \multirow{2}{18mm}{Repetitions}  \\ 
    \cline{3-6}
              &  & Total   & Training  & Validation    & Test &   \\
    \hline
         1-person   &  within subject    &   19280       &   13496       & 2892    & 2892  & 2         \\
         3-persons  &  across-subjects   &   57840       &   40488       & 8676    & 8676  & 10                  \\
         5-persons  &  across-subjects   &   96400       &   67480       & 14460   & 14460 & 10                 \\
         7-persons  &  across-subjects   &   134960      &   94472       & 20244   & 20244 & 10                 \\
         9-persons  &  across-subjects   &   177570      &   125514      & 26028   & 26028 & 10                  \\
    \hline
    \end{tabular}
    \caption{Details of within and across-subjects models with number of training, validation and test instances, as well as the number of Monte Carlo repetitions.}
    \label{tab:trainingDetails}
\end{table*}

Several models are trained within and across subjects to test the research hypothesis, as listed in table \ref{tab:trainingDetails}. 
Since each participant watched $40$ videos, then the number of total sequences associated with each participant equates to $19280$ ($482x40$). 
The canonical approach employed in machine learning to create generalisable models would be to shuffle these sequences and split them into training, validation and test sets.
However, although technically valid, performing such a shuffle for training a within-subject model would generate a training set that will likely contain some sequence from each video. In other words, each video would have a certain amount of representative data in the training, validation and test sets. To further increase generalisability, it is decided that the training set contains entire data from random $70\%$ of the possible videos, and the validation and test sets, respectively $15\%$ of the data associated with the remaining videos. 
Thus the shuffle is done at the video level, and data associated with $28$ random videos are selected as the training set ($482x28=13496$ training sequences), data from $6$ different random videos for the validation set ($482x6=2892$ training sequences), and the data from the remaining videos for the test set. In this way, the generalisability is exploited across unseen test videos, expected to lead to different cognitive load fluctuations than those used for training and validating models. 
The same rationale is applied to across-specific models. The only difference is that the training, validation and test sets contain data from a random number of participants, as listed in table \ref{tab:trainingDetails}. In other words, for example, for a 3-persons model, $3$ splits are performed for each participant individually. Then the resulting individual training, validation and test sets are concatenated to produce larger sets.\\

$32$ within-subject CNN models (figure \ref{fig:cnn}) are trained for participants twice with different batch sizes ($32$ and $100$). This step aims to understand batch-size manipulation to validate and test errors. 
The rationale is to analyse the trade-off between generalisability and computational resource consumption since it is known that larger batches lead to better convergence to the global optima of the objective function but at the cost of slower convergence since more memory is requested and more computations are performed. Instead, smaller batches allow the model to start learning earlier,  before seeing all the data, with lower consumption of computational resources. Still, it is not guaranteed that the model converges to the global optima, thus with a negative impact on its generalisability.
After assessing the ideal batch size, across-subject models are trained with incremental complexity, in terms of a higher volume of data coming from an increasing number of participants, to assess whether their generalisability still holds with a higher heterogeneity in the EEG signals. 
Additionally, to reinforce the analysis, repeated Monte Carlo sampling is performed for each across-subject model, with a random selection of participants at each repetition. 
Table \ref{tab:trainingDetails} summarises the number of training, validation and test sequences used and the number of repetitions for each training configuration. The training dataset is not augmented in any way, for example, by employing image zooming or flipping techniques,  because of the distinct interpretations of direction and location in the EEG topographic-maps that correspond to specific cortical regions.
Training is conducted by optimising the Mean Squared Error (MSE) loss function:
$$
\frac{1}{n}\sum_{i=1}^{n}(y_i-\hat{y_i})^2
$$
with $n$ the number of sequences (of length $7$), $y_{i}$	the	observed brain rate for that sequence (in the 8th position) and 
$\hat{y}_{i}$ the predicted brain rate for that sequence.
Validation and test MSEs is monitored during and after training. Also, Mean Absolute Percentage Error (MAPE) is computed:
$$
\frac{100\%}{n}\sum_{t=1}^{n}\left |\frac{y_i-\hat{y_i}} {y_i}\right|
$$
where $y_{i}$ is the observed brain rate and $\hat{y_i}$ is the predicted one. Their difference is divided by the actual observed brain rate $y_{i}$. The absolute value in this ratio is summed for every predicted brain rate and divided by the number of sequences $n$. MAPE comes under percentage errors and it has been selected because these errors are scale independent, thus especially suitable for across-subject models and because it is easier to interpret and explain.
As mentioned earlier, the parallel CNNs share weights, thus potentially producing different gradients in different internal layers. As a consequence, a smaller learning rate, set to $1e-3$, was employed when applying the Stochastic Gradient Descent (SFD) to the CNNs. 
Similarly, the whole convolutional-recurrent neural network was trained with a small learning rate of $1e-4$ optimised with the Adam
algorithm \cite{kingma2015amsterdam}, shown to
achieve reasonable fast convergence rates,  with decay rates of first and second moments set to $0.9$ and $0.999$ respectively.\\

The overall final neural network devised contains a large number of parameters (1.62 million) and considering that a different number of models are trained with an increasing amount of training instances per model, with each instance being a tensor of $32x32x5x7$ (where $32x32$ is the size of the spatial-preserving topographic maps, $5$ is the number of EEG bands, $7$ is the number of EEG windows, that means the length of the trainable sequence), a significant demand on computational resources, in terms of memory and processing power, is required. 
Additionally, many parameters can make each trained model susceptible to overfitting. 
Therefore, several measures are taken into account.  
As mentioned earlier, all the CNN networks share parameters across the $7$ frames. Thus a good amount of parameters in the overall architecture were removed. 
Dropout layers were added after each fully connected layer, with a probability of $0.5$ to minimise overfitting \cite{hinton2012improving,krizhevsky2012imagenet}.
Similarly, an early stopping training mechanism is employed to avoid training models when it is no longer necessary, thus saving a significant amount of time. 
This is an optimization procedure that is also used to minimise overfitting without compromising on model accuracy.
In detail, this is a regularization technique that stops training when the updates of the model's parameters no longer yield improvement on a validation set after consecutive $E$ epochs. 
The value $E$ is called patience, and in this study it was set to $6$, after some trials. This means that the training phase early stops automatically when the error associated with the validation set does not reach a lower value for $6$ consecutive epochs, and the $Eth$-last model is retained as the final model.\\

Data up to $9$ people are considered to train a single across-subject model since this is the maximum amount of data that the selected machine has been estimated to process with its resources. 
In particular, this machine is an Alienware Aurora R8 (model: 02XRCM), Intel Core i7-8700 (6-core, 12 threads), 64 bits, 12Mb L2-cache, 32GB DDR-SRAM, 2 additional graphics cards (GeForce RTX 2070), with the Linux Mint 19.2 operating system, and an internal local total storage of 4 TeraBytes, comprising a primary 1TB SSD (Solid State Drive) hard-disk (model: SK Hynix, PC601 NVMe), a 3.5-inch 2TB hard-drive (model: Seagate BarraCuda ST2000DM008-2FR102) and an additional 1TB SSD hard-disk (model: 2-Power SSD2044A). For allowing training of across-specific models (up to 9 persons), a Swap RAM  of 0.5TB was created.

\section{Results} \label{sec:results}
\begin{figure}[ht]
\centering
    \includegraphics[scale=0.55]{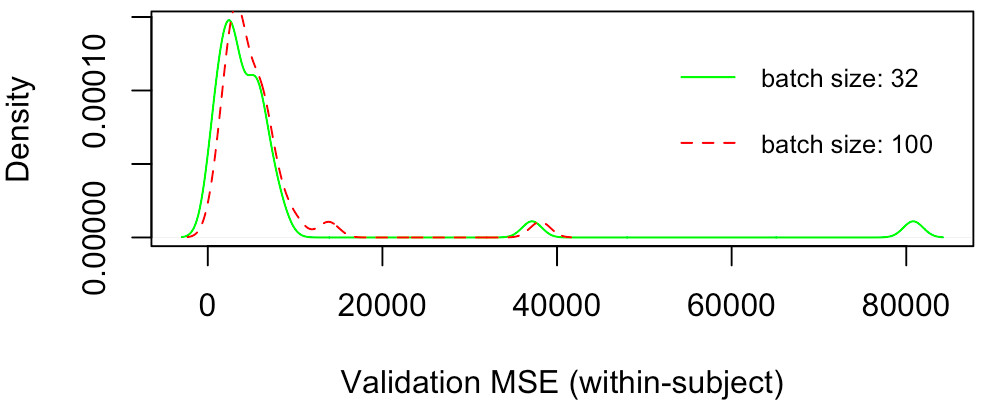}
    \includegraphics[scale=0.55]{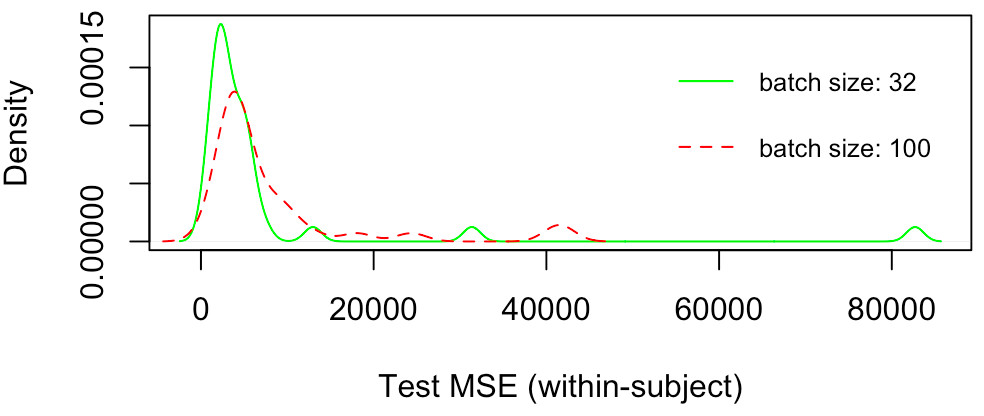}
    \caption{Comparison of validation and test Mean Squared Error for within-subjects CNN models trained respectively with batch size of dimension 32 and 100.}
    \label{fig:batch_size_comparison}
\end{figure}

\begin{figure}[ht]
    \centering
    \includegraphics[scale=0.54]{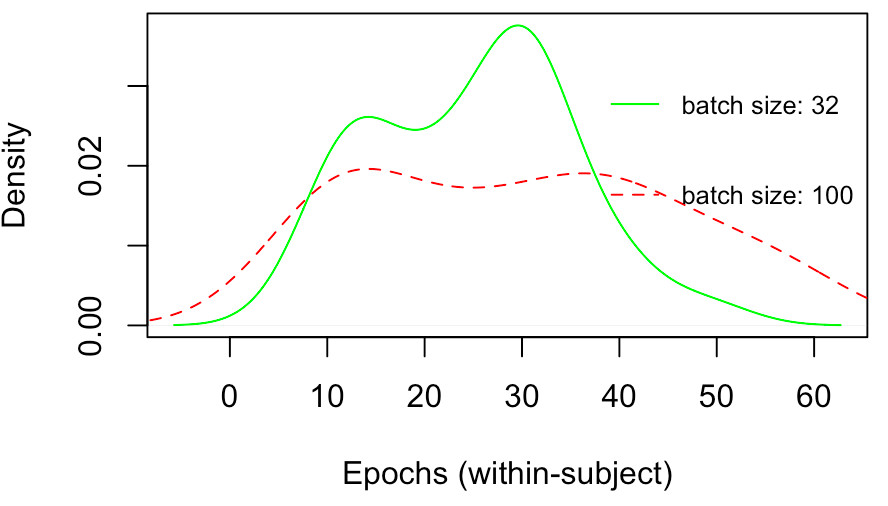}
    \caption{Comparison of the number of epochs to train the  within-subjects CNN models respectively with batch size of dimension 32 and 100.}
    \label{fig:cnnEpochs}
\end{figure}

Figure \ref{fig:batch_size_comparison} depicts the density plots of the validation and test mean squared errors (MSEs) for the $32$ within-subject models trained only by employing the CNN architecture (\ref{fig:cnn}), respectively with batch size of $32$ and $100$.
Similarly, figure \ref{fig:cnnEpochs} depict the density plots of the number of epochs necessary to train the within-subject CNN architectures, respectively, with a batch size of $32$ and $100$, with a minimum of $7$ epochs to a maximum of $60$.
No significant difference exists in the validation and test errors, with the batch size of $32$ leading to slightly better (lower) MSEs. However, although not significantly different, on average, the number of epochs necessary to train CNN models with batch size $32$ is lower than that associated with batch size $100$.
Every epoch for the within-subject model, with the current machine, required on average $300$ seconds (5 minutes), thus, the finalisation of training, according to the minimum and a maximum number of epochs ($7$ and $60$), required between $2100$ to $18000$ seconds (35 and 300 minutes).
Therefore, $32$ was the batch size selected for training the subsequent within-subject and across-subject models with the full architecture (figure \ref{fig:neuralnetwork}) since it leads to a lower number of training sequences in one forward/backwards pass, thus lower consumption of memory, as well as a lower number of training epochs, saving a great amount of time.

\begin{figure}[ht]
    \centering
    \includegraphics[scale=0.39]{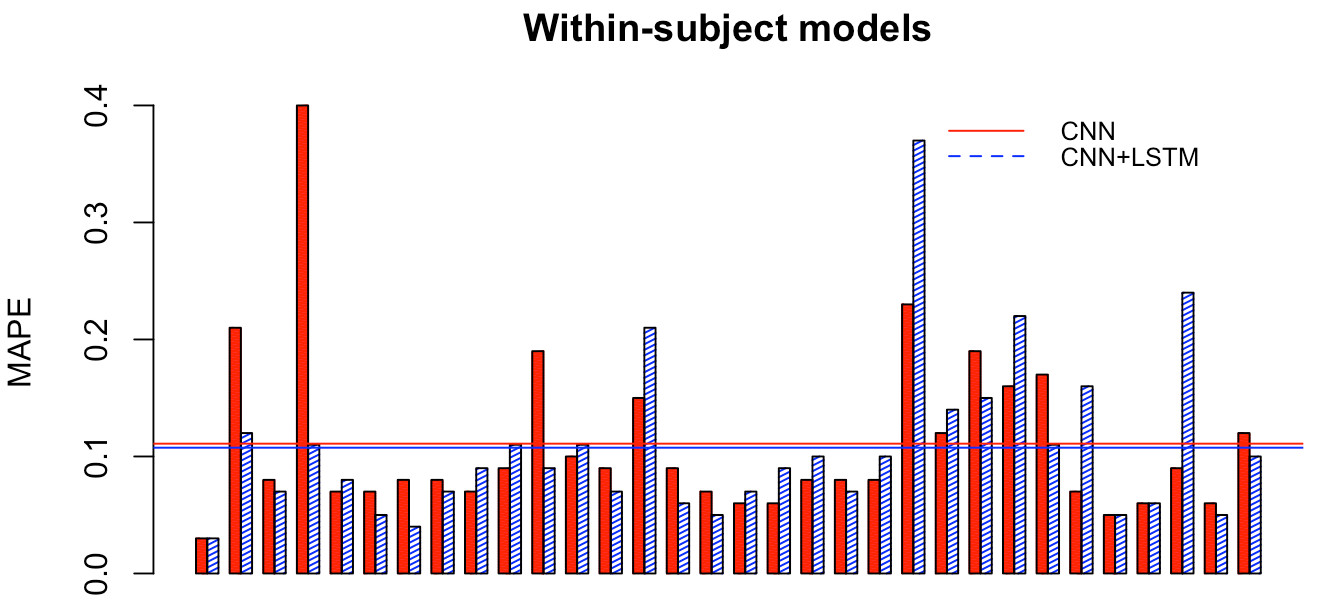}
    \caption{Paired histogram of the Mean Absolute Percentage Errors (MAPE) of the test data of the 32 within-subject models respectively trained only with the single Convolutional Neural Network (CNN), and the Convolutional/Recurrent Neural network (CNN+LSTM).}
    \label{fig:mape_within}
\end{figure}
\begin{figure}[ht]
    \centering
    \includegraphics[scale=0.52]{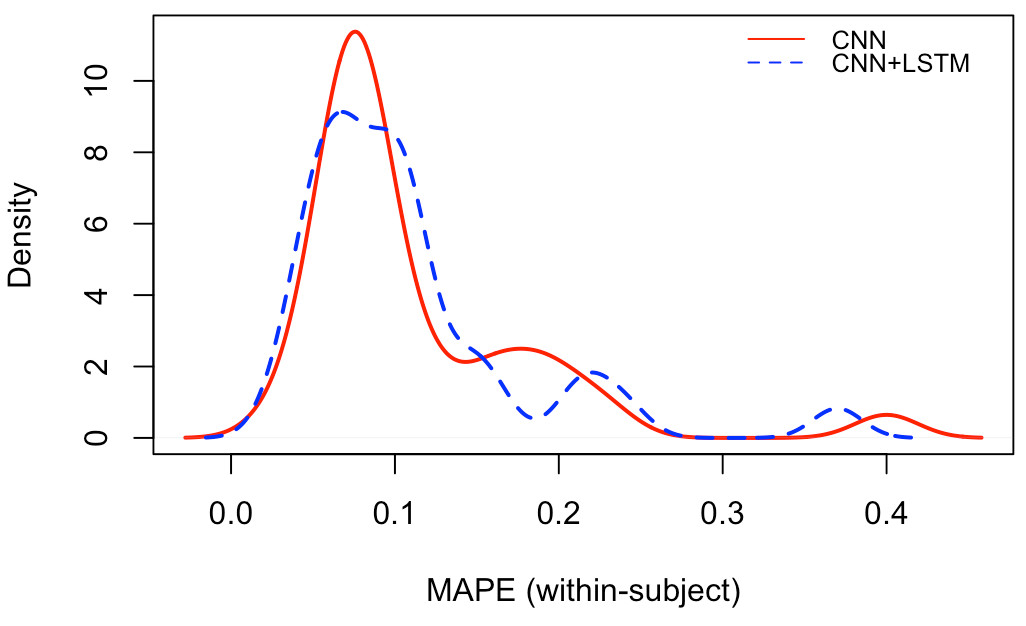}
    \caption{Density plot of the Mean Absolute Percentage Errors (MAPE) of the test data of the 32 within-subject models respectively trained only with the single Convolutional Neural Network (CNN), and the Convolutional/Recurrent Neural network (CNN+LSTM).}
    \label{fig:mape_within_density}
\end{figure}

\begin{figure*}
\centering
    \includegraphics[scale=0.35]{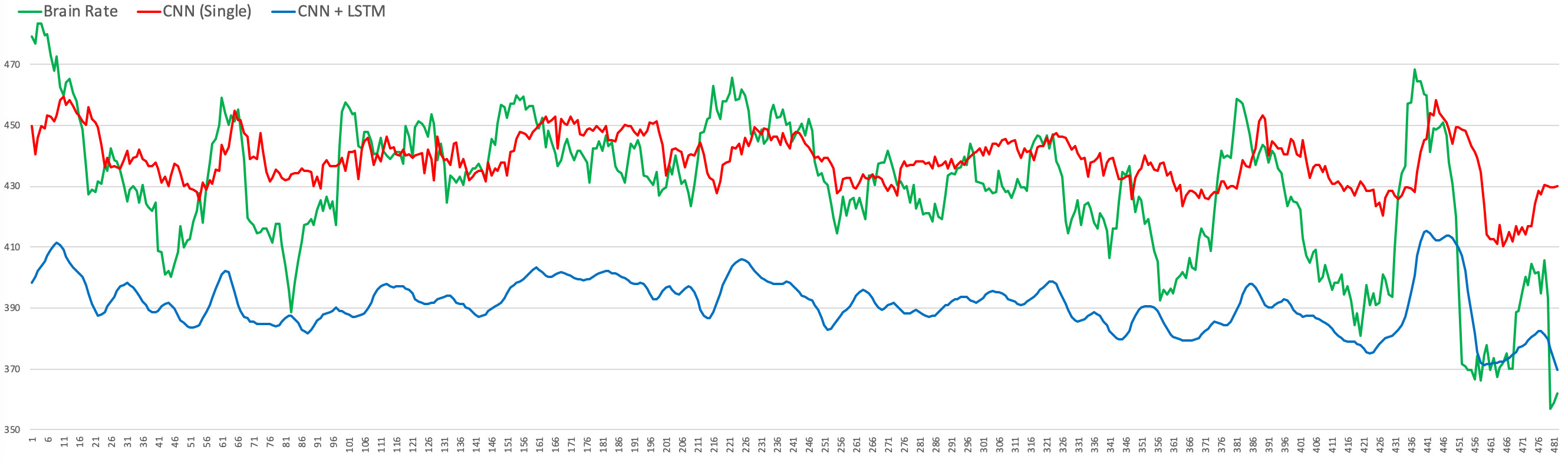}
    \includegraphics[scale=0.35]{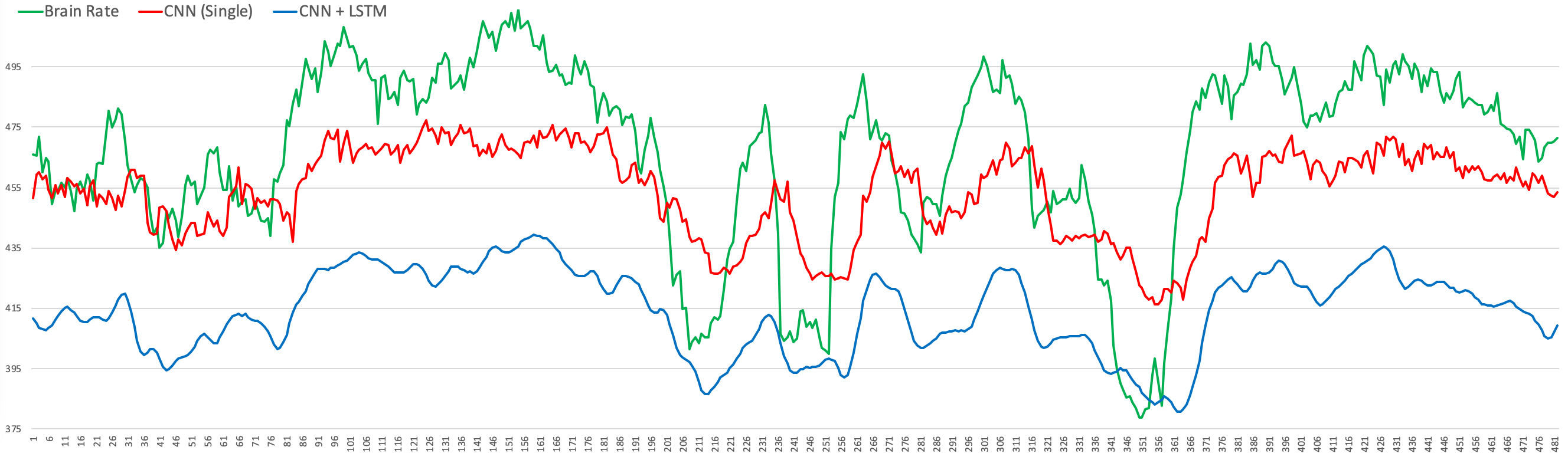}
    \caption{Illustrative comparisons of the brain rate index, the single Convolutional Neural Network (CNN) predictions and the Convolutional/Recurrent Neural Network (CNN+LSTM) predictions for two random participants and a random video used in the test set.}
    \label{fig:exampleComparisonsPreds}
\end{figure*}

Figure \ref{fig:mape_within} and \ref{fig:mape_within_density} depict the Mean Absolute Percentage Errors (MAPE) for the test data of the within-subject models for the $32$ participants, trained first with the single CNN architecture of figure \ref{fig:cnn} for learning the weights (in full red), and with the convolutional/recurrent neural network with the parallel CNNs, sharing such weights, and the LSTM component for temporal learning (figure \ref{fig:neuralnetwork}) (in dashed blue).
As it is possible to notice, the test MAPE has mean $0.111$ (Std: $0.073$) for the single CNN models and mean $10.75$ (Std: $0.070$) for the CNN+LSTM models.
These results demonstrate that the brain rate prediction for each participant's unseen test data is good because the forecast is only off by roughly $10\%$. 
However, at first glance, it seems that the impact of the addition of the recurrent component (the Long Short Term Memory), as in the architecture depicted in figure \ref{fig:neuralnetwork}, does not add much value in minimising the MAPE. 
This seems to point to the individual capability of the single CNN architecture (figure \ref{fig:cnn}) to learn the relevant patterns, intricacies and relationships in the data in the shape of topographic head maps containing information in the 5 EEG frequency bands for the specific window length used (2 seconds).
However, the LSTM layer takes a sequence of $7$ outputs from the single CNNs (in addition to a vector containing their variational information) and tries to fit the brain rate associated with the next window (the $8th$ after the sequence).
The fact that the MAPE of the CNN+LSTM does not significantly change (decrease) does not mean that the LSTM did not learn any temporal relationship and dependency in the input sequences. 
This can be demonstrated by inspecting figure \ref{fig:exampleComparisonsPreds}, whereby the brain rate index, the predictions of the single CNN  model and those of the CNN+LSTM for some within-subject models associated with random participants and a random video in their respective test sets, are compared.
In detail, these figures show that the brain rates (green), computed for each of the 482 instances, as explained in section \ref{sec:training} (and depicted in figure \ref{fig:pipelineTraining}), associated with a specific video that a participant has watched, not used for training the within-subject model of that participant, are reasonably approximated by the single-CNN within-subject model (red). 
However, the brain rate indexes seem better approximated by the CNN+LSTM within-subject model (blue). \\

The comparisons of figure \ref{fig:exampleComparisonsPreds} highlight a number of things. 
Firstly, the main bursts in the brain rates are also grasped by the CNN and the CNN+LSTM models.
However, those associated with the CNN (red) are shifted a bit to the right (x time axis) when compared to those associated with the CNN+LSTM (blue), which seem to be more aligned to the brain rates (green) over time.
This is confirmed by the  Person correlation coefficient, which on average for participants and testing videos, is $0.5$ for the CNN models and $0.7$ for the CNN+LSTM models.
This means that the LSTM layer in the CNN+LSTM architecture did learn some temporal relationships and long/short-term dependencies. 
The CNN+LSTM predictions are smoother than those produced by the single CNN, and this might be justified by the fact that they are based on the information taken from the precedent 7 consecutive EEG windows over time. For the same reasons, this might be the reason why the scale (y-axis) of the predictions of the CNN+LSTM (blue)  is a bit lower than the others (blue and green).\\

\begin{figure}[ht]
    \includegraphics[scale=0.365]{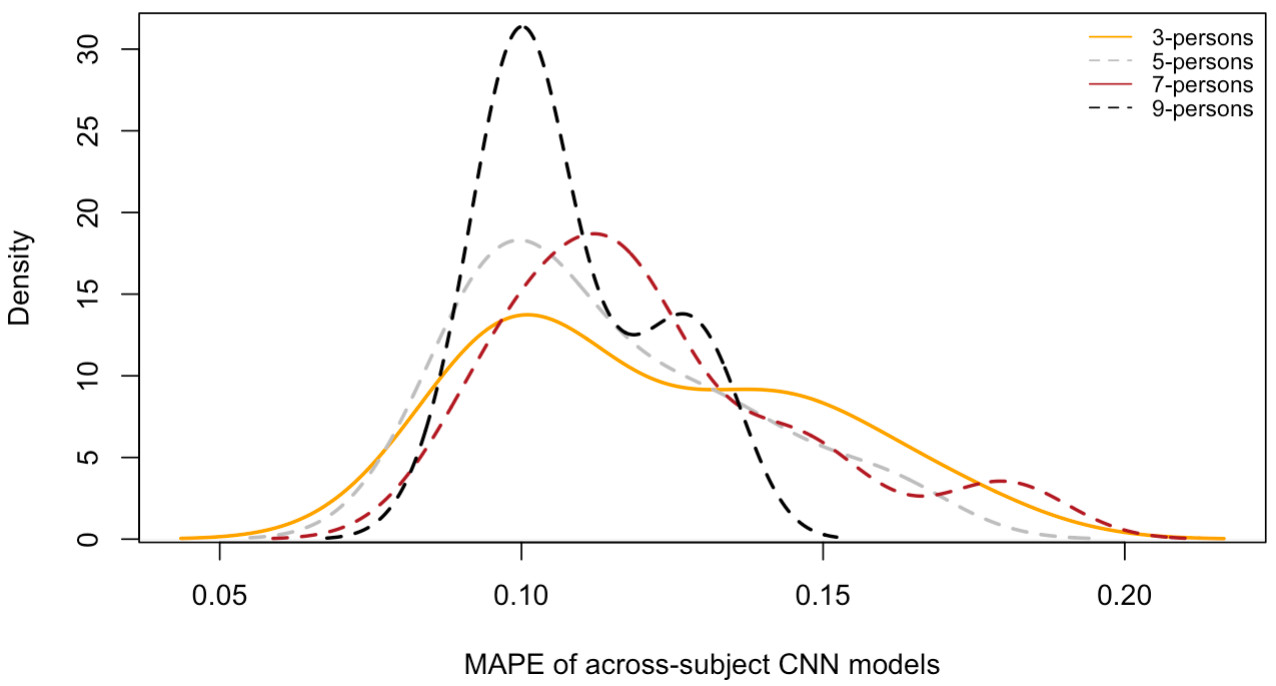}
    \includegraphics[scale=0.365]{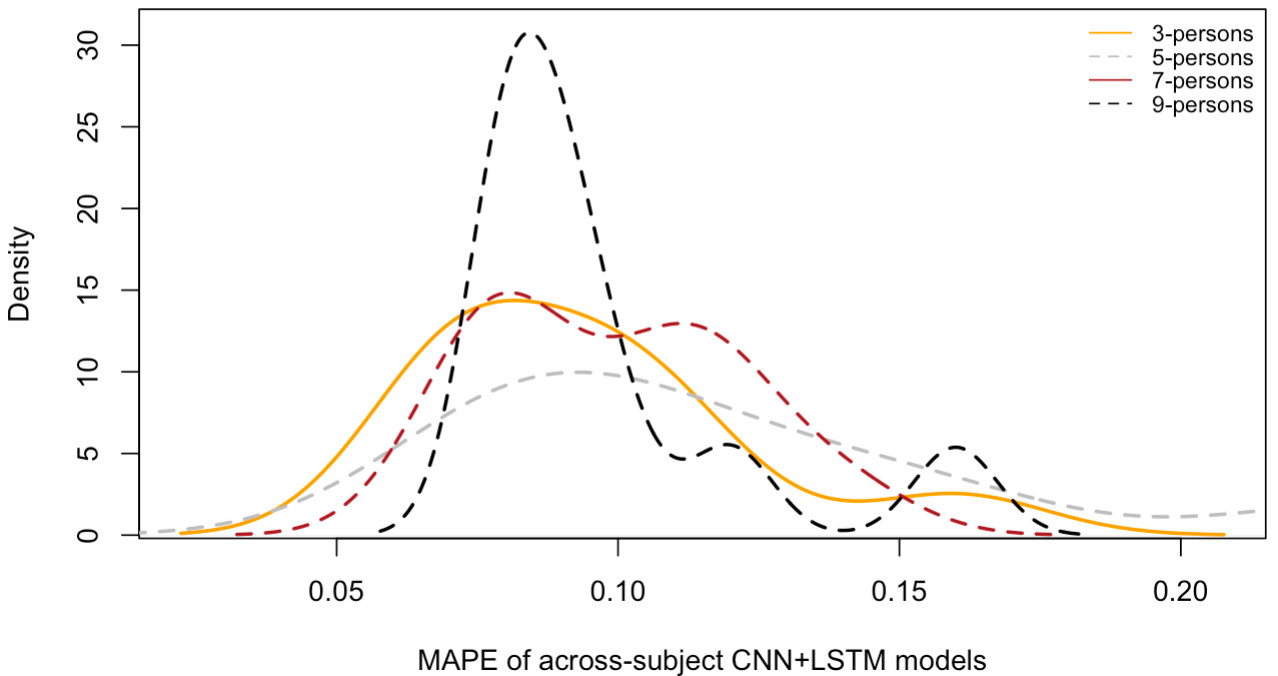}
    \caption{Comparisons of the test Mean Absolute Percentage Error (MAPE) of the across-subject models grouped by the type of architecture which is the single convolutional neural network (CNN) and the convolutional/recurrent neural network (CNN+LSTM).}
    \label{fig:compMAPE_CNN_CNNLSTM_across-subject}
\end{figure}

Regarding the across-subjects models, as planned in table \ref{tab:trainingDetails}, figure \ref{fig:compMAPE_CNN_CNNLSTM_across-subject} depicts the density plots of their Mean Absolute Percentage Errors (MAPEs) on the test sets. 
In detail, each density curve contains the MAPEs associated with the test sets of 10 models, each trained with the respective number of random people.
As it is possible to see, the test MAPEs are lower on average for those models trained with material taken from 10 people (black), followed by those trained with 7 (brown), 5 (grey) and 3 people (yellow). 
Additionally, the standard deviations (width of each curve) are smaller (thinner) for those trained with data from more people and larger for those trained with data from fewer people.
This means that smaller standard deviations are associated with more steady models because these are capable of predicting brain rates on the test data more consistently.
These results might seem intuitive because it can be argued that the more training material, the higher capacity a model has to learn.
However, training material comes from different numbers of people, selected randomly at each run, and their cerebral responses are different while watching videos, exhibiting different power activations and temporal dynamics. 
This introduces a higher variability within data, thus making a model prone to confusion while learning. 
Despite this, across-subject models can mitigate the influence of such an increasingly higher variability and can learn consistent higher-level representations that are more generalisable across people. \\

\begin{figure}[ht]
    \centering

    \includegraphics[scale=0.43]{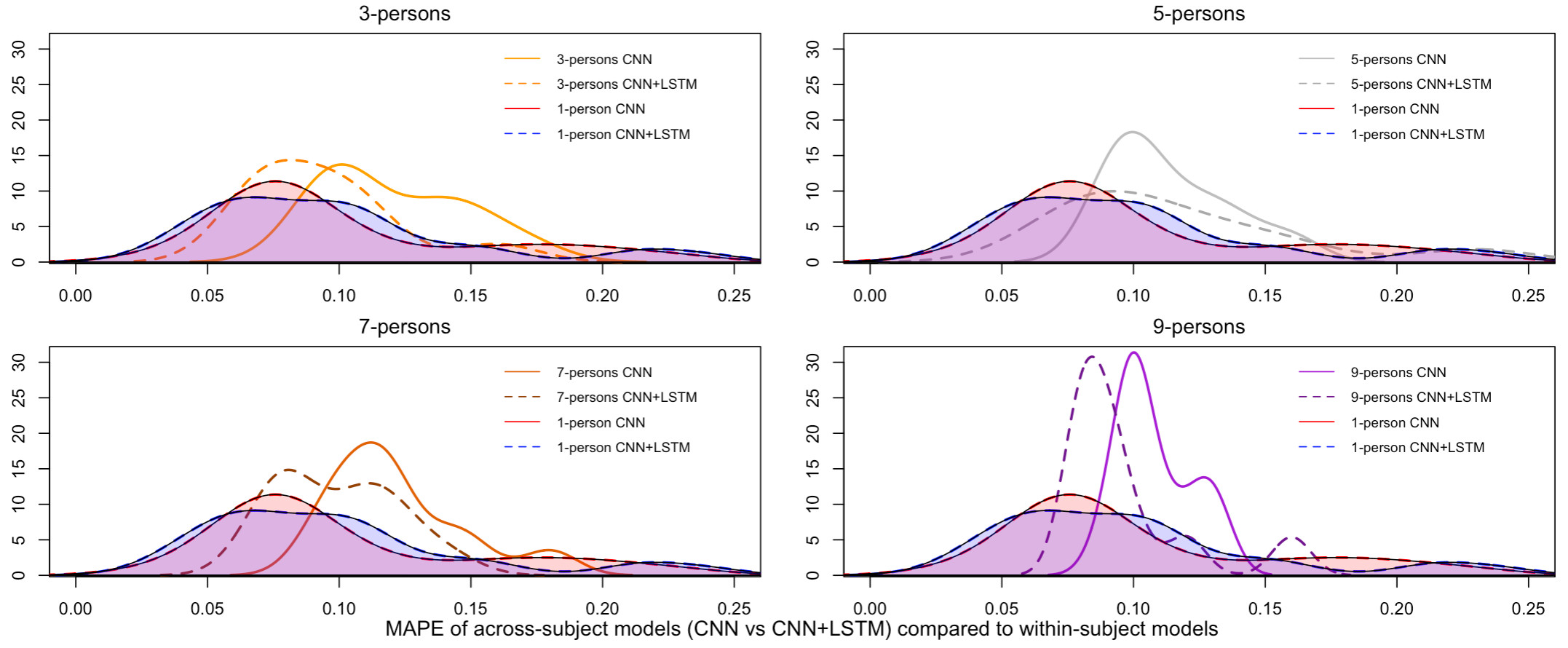}

    \caption{Pairwise comparisons of the test Mean Absolute Percentage Error (MAPE) of the across-subject models trained respectively with the single convolutional neural network (CNN) and the convolutional/recurrent neural network (CNN+LSTM) compared to the within-subject models.}
    \label{fig:pairwisecompMAPE_CNN_CNNLSTM_across-subject}
\end{figure}
 
Figure \ref{fig:pairwisecompMAPE_CNN_CNNLSTM_across-subject} plots the pair-wise comparison of the across-subject models trained with the single CNN and the CNN+LSTM architectures, grouped by the number of people, and the density curve associated to the MAPEs of the within-subject models, used here as baseline. 
Noticeably, the density plots associated with those models trained with the CNN+LSTM architecture (dashed lines) contain lower MAPEs on the test sets than those associated with the models trained with the CNN only (continuous lines).
This means that the addition of the Long-Short Term Memory (LSTM) layer for temporal learning had an impact on building more accurate models, although, in this study, not statistically significant. 
Additionally, these results suggest that the convolution of the topology-preserving topographic maps over space (down-sampling) could learn some repetitive high-level patterns within an EEG window (as set to 2 seconds). 
In other words, as expected in the research hypothesis set in section \ref{sec:design}, within-subject and across-subjects models can be induced from spatio-temporal spectral topology-preserving head maps derived from multi-channel EEG data to fit a brain rate, an index of cognitive activation, with low error-rates, demonstrating the existence of recurrent patterns of cognitive load over time. A more detailed interpretation of such results, along with a discussion of the strengths and limitation of the designed method for cognitive load modeling, is done in the following section.

\section{Discussion} \label{sec:discussion}
The computational method described and tested in the previous sections is fully automated and allows the induction of a model of cognitive load from EEG data based on deep learning without requiring human intervention. In summary, this novel method:
\begin{itemize}
    \item is based on data-driven deep-learning techniques for automatic inductive learning  \cite{lecun2015deep};
    \item is built upon electroencephalography (EEG), a non-invasive method for gathering brain responses with high-temporal resolution \cite{craik2019deep};
    \item is sensitive to brain responses variation over time thanks to its recurrent neural network component \cite{hochreiter1997long};
    \item is robust to deformation and translation of signal in space, and frequency,  thanks to the ability of its convolutional neural network component to learn meaningful representations \cite{lecun1998gradient};
    \item is built upon 2D spectral topology-preserving head maps that are rich in information and also more explainable than vectorial data \cite{LongoGLKH20, VILONE202189, Vilone2021Output};
    \item is self-supervised and does not require human intervention and explicit declarative knowledge \cite{banville2021uncovering};
    \item is constructed upon a brain rate, a measure of cognitive activation, and treated as an index of cognitive load that considers cortical brain oscillations weighted over the potentials of all the canonical EEG bands;
    \item is flexible with short EEG segments, thanks to its time-slicing procedure over cortical recordings;
    \item it is adjustable and customisable because it can be trained on EEG data collected from a variable number of electrodes, it can be employed with different ranges for the five EEG bands (delta, theta, alpha, beta, gamma), and with EEG windows of varying size;
    \item it is replicable and open to falsifiability \cite{popper2005logic}, supporting the formation of models of cognitive load with higher generalisability.
\end{itemize}

This method allowed the fully-automated formation of within-subject and across-subject models of cognitive load from EEG signals. 
These models fit a brain rate, an index of cognitive activation, with good accuracy, measured by the Mean Absolute Percentage Error (MAPE) on the test sets, demonstrating a good degree of generalisability to unseen data.
In detail, each within-subject model, trained with EEG  material from a single person, could predict the brain rate of unseen EEG data - as encoded with spatially preserving topographic head-maps built upon 32 channels - with a MAPE of 0.11 and 10.75 (std 0.073, 0.070), only using a convolutional neural network architecture for spatial learning, and its extension with a long-short term memory layer for temporal learning, respectively.
The across-subject specific models, induced from an increasingly higher amount of EEG data from different people, confirmed these results and maintained the same testing accuracy as measured with MAPE, despite the increasing variability within training data.
This perseveration in achieving similar testing accuracy, despite a higher  variability in training data, can be seen as positive because it highlights the existence of some patterns within EEG data that are repetitive and stable.
This observation might be linked to microstate theory which assumes that distributions of activity across the scalp persist for milliseconds before changing into a different pattern \cite{MICHEL2018577}. 
EEG microstates can be seen as transient, quasi-stable patterns of an electroencephalogram  \cite{wackermann1993adaptive,khanna2015microstates}.
An analogy can be applied to the findings obtained in this current work, and the trained models might have learned quasi-stable patterns of mental activation fluctuations, as modelled with a brain rate.
The convolution applied to the spatially preserving topographic head-maps, built over five EEG frequency bands, has already led to the development of within and across-subject models with good accuracy. This means quasi-stable high-level representations might be induced from the convolutional operations that can be successfully mapped to a brain rate.
Also, this view might be enforced by the minimal decrement of the test MAPEs obtained by those models trained with the LSTM layer in the neural network for temporal learning. The fact that it was minimal suggests that the sequence of convoluted representations over time is not as important as the actual representations alone, taken individually, which seem to be already rich in information and able to learn certain repetitive patterns of cognitive activation.

\section{Conclusion}
Cognitive Load, often referred to as Mental workload \cite{HancockLongo2021}, is one of the most invoked concepts in the disciplines  of human factors, with important utility within human-computer interaction, neuroscience and education \cite{longo2022human}.
Unfortunately, a reliable, generally applicable computational method for cognitive load modelling does not exist yet, complicating applied research.
This research, the first of its kind, was aimed at developing a method for cognitive load modelling with generalisability in mind, supporting its application across disciplines,  replicability, comparisons across studies and thus enabling falsifiability.
All these advantages are aimed at supporting research on cognitive load modelling at a larger level,  avoiding the creation of another ad-hoc, field-dependent, knowledge-dependent and application-driven method of mental workload that has little chance of being generally applicable across empirical works.
This novel method employs Deep Learning techniques of Artificial Intelligence, for the automatic formation of models of cognitive load, in a fully unsupervised way, drastically limiting human intervention and declarative knowledge. 
These models work on continuous EEG data, thus having a great temporal resolution.
They are built upon a newly designed notion of brain rate, a particular index of cognitive load derived from the five EEG frequency bands (delta, theta, alpha, beta, gamma).
This method works on spatially-preserving topographic head-maps of cognitive activation, offering spatial resolution and supporting diagnosticity. 
In this study, these maps are based on spectral information derived from the five EEG bands, which are known to be rich in information for deriving mental states and facilitating the analysis and interpretation of human behaviours.\\

Findings suggest that within-subject and across-subjects models of cognitive load, developed with the newly devised computational method, are accurate enough, exhibiting a low prediction error on unseen data, thus showing a good degree of generalisability. 
They suggest that certain high-level representations from EEG data in the frequency bands can be extracted automatically, frequently appearing over time. 
However, these existing repetitive blocks of mental activation do not seem to be repetitive over time, in line with the non-stationary nature of brain activation.
In other words, frequent, quasi-stable high-level representations of cognitive activation exist, but these are not repetitive over time. 
Additionally, these representations seem to be repetitive across-subjects, with important implications for the research field of mental workload.
Their existence might suggest that general patterns of cognitive load exist, and these are subject-independent, therefore having a great generalisability.
However, to confirm this claim, further studies are needed.\\

Future work will include replicating the method developed in this research study with varying time window sizes and investigating how these influence the accuracy of resulting cognitive load models.
A layer of interpretability for the automatically extracted higher-level representations will be deployed, in line with principles and practices from Explainable Artificial Intelligence \cite{VILONE202189,Vilone2021Output} and knowledge-representation \cite{LONGO2021106514,LONGO2021106514}. This will help understand the shape of these representations, and the recurrent activated brain regions, giving analysts a richer level of interpretability. It will also serve as a layer of explainability, providing analysts with tools for explaining spatial and temporal dynamic of cognitive activation. 
The inferences of these models of cognitive load can be compared against other indexes such as the theta-to-alpha or alpha-to-theta band ratios \cite{RaufiLongo2022}, increasing their meaningfulness and validity. 
Eventually, studies can be devoted to the development of additional recurrent neural networks for understanding the temporal aspects of the high-level representations of cognitive activation, and establishing if there exist sequences, and their lengths, that are repetitive and recurrent over time. 
These future avenues will expand the science of mental workload and support the formation of models of cognitive activation with an increasing accuracy and generalisability, in turn facilitating the analysis of human behaviour.

\bibliographystyle{plain}
\bibliography{main}

\end{document}